\documentclass[twocolumn,showpacs,preprintnumbers,amsmath,amssymb]{revtex4}
\usepackage{graphicx}
\usepackage{dcolumn}
\usepackage{bm}
\usepackage{epsfig}

\newcommand \bea{\begin{eqnarray}}
\newcommand \eea{\end{eqnarray}}
\newcommand \ga{\raisebox{-.5ex}{$\stackrel{>}{\sim}$}}
\newcommand \la{\raisebox{-.5ex}{$\stackrel{<}{\sim}$}}

\begin{document}
\title{Sound modes at the BCS-BEC crossover}
\author{H. Heiselberg}

\altaffiliation[Also at ]{Univ. of S.Denmark}
\email{hh@ddre.dk}
\affiliation{Danish Defense Research Establishment, Ryvangsalle' 1, 
DK-2100 Copenhagen \O, Denmark}

\begin{abstract}
First and second sound speeds are calculated for a uniform superfluid gas
of fermi atoms as a function of temperature, density and interaction strength. 
The second sound speed is of particular interest as it is a clear signal
of a superfluid component and it determines the critical temperature.
The sound modes and their dependence on density, scattering length and 
temperature are calculated in the BCS, molecular BEC and unitarity limits and a
smooth crossover is extrapolated. 
It is found that first and second sound undergo 
avoided crossing on the BEC side due to mixing. Consequently, they are
detectable at crossover both as density and thermal waves in traps.
\\ 

\end{abstract}
\maketitle

\section{Introduction}

Recent experiments probe systems of strongly interacting Fermi and
Bose atoms and molecular bosons at low temperatures. Evidence
for a continuous crossover between BCS to a molecular BEC superfluid 
is found by measurements of
expansion\cite{Thomas,Regal,Bourdel,Ketterle,Jochim}, collective modes
\cite{Kinast,Bartenstein}, RF spectroscopy \cite{Chin}, thermal
energies and entropies \cite{entropy}, and correlations
\cite{Greiner}.  The next generation of experiments will measure sound
velocities in these systems. 
Second sound has been studied in He$^3$,  He$^4$ 
and BCS superfluids (see, e.g., \cite{Vollhardt},\cite{FW}).
Second sound is particularly interesting
as it is a clear signal of a superfluid component and determines the
critical temperature \cite{Ho}. The dependence of the sound speeds on
temperature, density and interaction strength will constrain the
equation of state and models.

Eagles \cite{Eagles} and Leggett \cite{Leggett} described the smooth
crossover at zero temperature by a mean field gap equation. Extensions
to finite temperature especially around the superfluid critical
temperature can be found in
\cite{NSR,Haussmann,Randeria,Levin,Strinati,Griffin,Holland,Melwyn}. The
various crossover models differ in the inclusion of selfenergies,
whether one or two channels are employed, etc. Recently, the crossover
has been calculated in detail by quantum Monte Carlo (QMC)
\cite{Carlson,Casulleras} at zero temperature.  The experiments are
generally compatible with most these calculations and confirm
universal behavior \cite{Ho,HH} in the unitarity limit as well as
superfluidity at high temperatures 
\cite{Levin,Holland,HH,Timmermans,Stoof}.

The purpose of this work is to calculate sound velocities in the
superfluid phase and demonstrate that first and second sound undergoes
avoided crossing at the BEC-BCS crossover in uniform systems of strongly
interacting Fermi atoms. Second sound reveals the presence of a
superfluid and its critical temperature. The sound modes provide
details about the equation of state at the BEC-BCS crossover and
constrain crossover models.

The two-body interaction range is assumed to be short as compared to
the s-wave scattering length $a$ and the interparticle spacing or
$k_F^{-1}$, where $n=k_F^3/3\pi^2$ is the density of the Fermi atoms
with two equally populated spin states. The physics can be
expressed in terms the variable $x=1/(ak_F)$, which varies from
$-\infty$ in the dilute BCS limit through the unitarity limit $x=0$ at
the Feshbach resonance to $x\to +\infty$ in the molecular BEC
limit. 

The sound modes may be measured in
experiments as density waves in the expanding clouds. Density or
thermal perturbations in the trapped clouds may be introduced in the
center of the cloud, for example, by a pulsed laser before expansion,
or by other means \cite{Thomaspriv}. It is important that the two
sound modes mix as will shown below because both modes will be excited
by either a density or thermal perturbation. As the sound modes travel
through the cloud the density and temperature may vary.  It may
therefore be nontrivial to extract the sound speeds as function of
density, temperature and scattering length, and useful to have a model
calculation of both sound speeds.

The manuscript is organized such that the sound modes are first 
described in general, and then calculated in the BEC, BCS and unitarity limits
respectively.
The three limits are then approximately matched assuming a smooth crossover.
Finally, summary and conclusions are given.

\section{Sound modes}

We shall in the following 
assume that the thermal excitations collide frequently and
therefore can be assumed in local thermal equilibrium. This is the
collisional limit where hydrodynamics apply. The transition from zero
to first sound as systems change from collisionless to collisional is
described in, e.g., \cite{BP} and the hydrodynamics of superfluids in
\cite{LL,PS}. For a Fermi liquid the collision time $\tau\propto T^{-2}$
can become large at low temperaqtures. 
The hydrodynamic limit is therefore limited to long
wavelength (or equivalently low frequency $\omega$) 
sound modes such that $\tau\omega\ll 1$. This can always be
achieved in a bulk system but not necessarily in a finite system where
collective frequencies $\omega$ are finite. However, it was found in
\cite{Kinast,Bartenstein,Hu} that axial and radial collective modes can be
described by hydrodynamics at temperatures well below the
superfluid transition.  Thus atoms in traps appear collisional in a
wide region around the unitarity limit. Yet the collisionless limit is
expected in traps at very low temperatures \cite{Smith}.

The presence of a normal and superfluid components leads to two sound
modes in the collisional limit
referred to as first and second sound. Their velocities $u_1$ and $u_2$ 
are given by the positive and negative solutions respectively of  
\cite{Zaremba,PS}
\bea \label{sss}
  u^2 = \frac{c_S^2+c_2^2}{2}\pm
       \sqrt{\left(\frac{c_S^2+c_2^2}{2}\right)^2-c_T^2c_2^2} \,.
\eea
The thermodynamic quantities entering are
the adiabatic $c_S^2=(\partial P/\partial \rho)_S$, the
isothermal $c_T^2=(\partial P/\partial \rho)_T$, and the
``thermal'' $c_2^2=\rho_s s^2T/\rho_n c_V$ sound speed squared.
The latter also acts as a coupling or mixing term.
The difference between the adiabatic and isothermal
sound speed squared can also be expressed as
\bea
  c_S^2-c_T^2=\left(\frac{\partial s}{\partial \rho}\right)^2_T 
           \frac{\rho^2T}{c_V}  \, .
\eea
Here, $\rho=\rho_n+\rho_s=nm$ 
are the total,  $\rho_n$ the normal and $\rho_s$ 
the superfluid mass densities, 
$s=S/\rho$ the entropy per unit mass and 
$c_V=T(\partial s/\partial T)_\rho$ the specific heat per unit mass.

The isothermal sound speed at zero temperature is
in both the hydrodynamic limit and for a superfluid gas given by
\bea \label{cT0}
   c_{T=0}^2 &=& \frac{n}{m} \left(\frac{\partial \mu}{\partial n}\right)_{T}
     \nonumber\\
   &=& \frac{1}{3}v_F^2 
  \left[ 1+\beta-\frac{3}{5}x\beta' +\frac{1}{10}x^2\beta''\right] \,,
\eea
where $v_F=\hbar k_F/m$, $\beta(x)=E_{int}/E_{kin}$ is the ratio of
interaction to kinetic energy,
$\beta'=d\beta/dx$, etc. QMC
calculations \cite{Carlson,Casulleras} find $\beta(x=0)=-0.57$ 
in the unitarity limit at zero temperature.
The continuous crossover found in
QMC and crossover models is confirmed experimentally for
the pressure, chemical potential, $\beta$, collective modes 
\cite{Thomas,Bourdel,Chin}
and recently also for the entropy \cite{entropy}.

 In the dilute BCS limit and at low temperature
$c_T=v_F\sqrt{(1+(2/\pi)ak_F)/3}$.  In the unitarity limit
$c_T=v_F\sqrt{(1+\beta(0))/3}\simeq 0.37v_F$.  In the dilute BEC limit
$c_T=\sqrt{(\pi/2)\hbar^2na_M/m^2}$, where $a_M$ is the molecular
scattering length.  According to QMC \cite{Casulleras} and four-body
calculations \cite{Petrov} $a_M\simeq 0.62a$, whereas in the Leggett and
several other crossover models $a_M=2a$.

We are mainly interested in the superfluid state at temperatures below
the critical temperature, which generally is less than that in the BEC
limit, $T\le T_c\la T_c^{BEC}=0.218E_F$, the temperatures will be much
smaller than the Fermi energy $E_F$ (in units where $k_B=1$). At such
low temperature the isothermal sound speed is given by Eq. (\ref{cT0})
in both the BCS and BEC limits. In the crossover model of Leggett the
isothermal sound speed is also to a good approximation given by the
zero temperature value for $T\la T_c$.  We shall therefore use
Eq. (\ref{cT0}) for $c_T^2$ in the following.

The entropy and mass densities have been calculated in the BEC
\cite{Zaremba,PS} and BCS \cite{sound} limits to leading order in
$ak_F$. One should, however, be careful in extrapolating to the
unitarity limit. For example a normal BEC is depleted as $a_M\to
+\infty$ and eventually quenched \cite{Cowell} whereas in crossover
models the condensate remains although it changes character from BEC
to BCS superfluid. Likewise, the collective mode frequencies in
trapped Fermi atoms increase for a normal BEC as the unitarity limit
is approached \cite{String}.  Crossover models \cite{Hu} and
experiments \cite{Kinast,Chin}, however, find the opposite for a
molecular BEC.  The BEC and BCS limits will now be described followed
up with a discussion on the extension and interpolation towards the
unitarity limit in various crossover models.

\begin{figure}
\begin{center}
\psfig{file=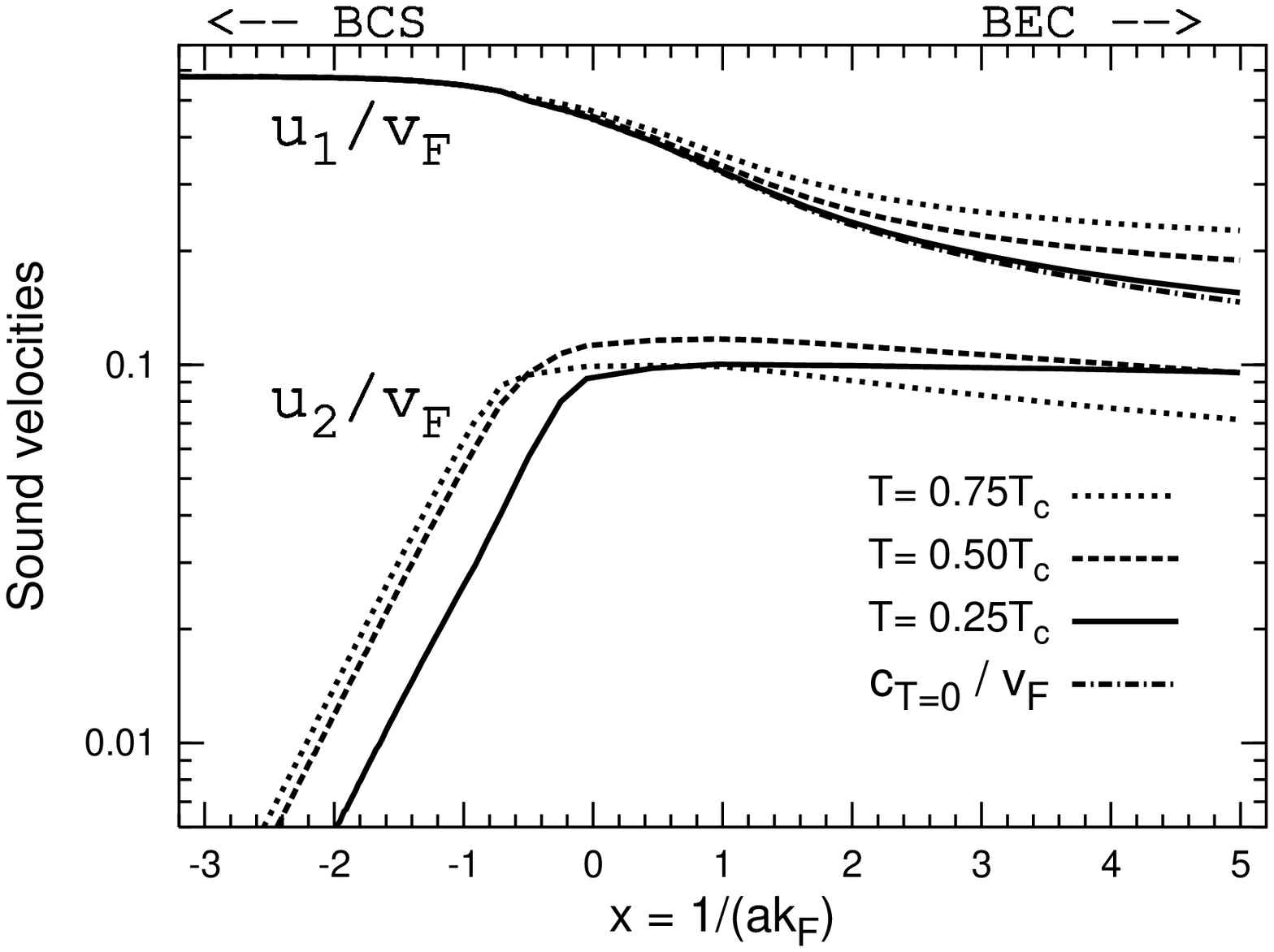,height=8.0cm,angle=-90}
\vspace{.2cm}
\begin{caption}
{First and second sound speeds for temperatures
$T/T_c=0.25,0.5,0.75$. The isothermal sound speed $c_T$ is
calculated from the Leggett 
crossover model. Sound speeds are calculated using entropies from Eq.
(\ref{SBEC}) on the BEC side, Eq. (\ref{OT}) on the BCS
side and from the measured entropy in the unitarity limit at $x=0$ (see text).
}
\end{caption}
\end{center}
\label{f1}
\end{figure}
\vspace{-.2cm}

\begin{figure}
\begin{center}
\psfig{file=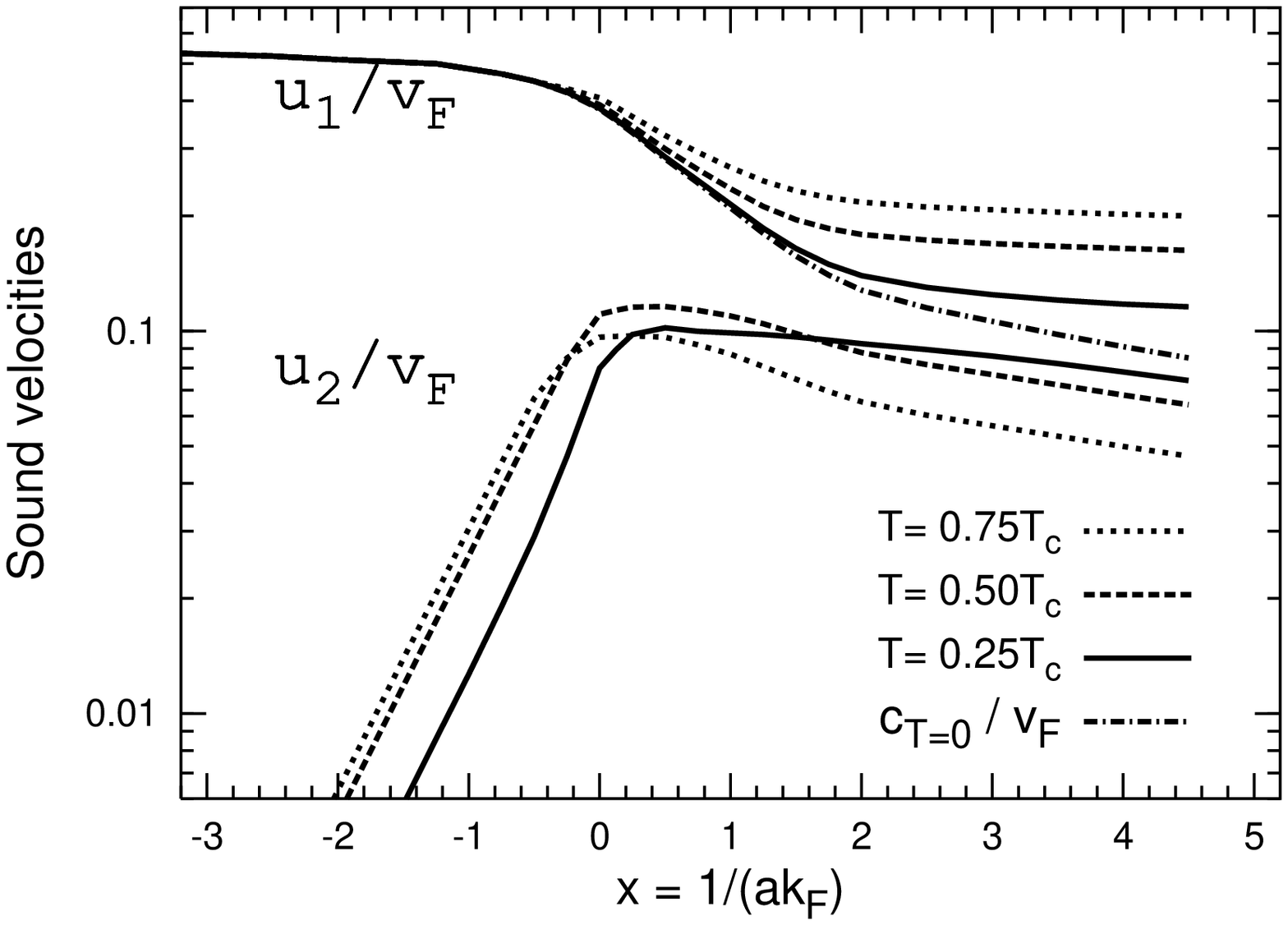,height=8.0cm,angle=-90}
\vspace{.2cm}
\begin{caption}
{As Fig.1 but with $c_T$ and $a_M=0.62a$ from QMC. Furthermore, 
induced interactions
are included on the BCS side which reduce $T_c$. 
}
\end{caption}
\end{center}
\label{f2}
\end{figure}
\vspace{-.2cm}

\section{BEC limit}
 
Crossover models and calculations find different
molecular scattering length in the BEC limit as mentioned above. Also
the effective mass $M$ in the dispersion relation $E_q=q^2/2M$ differs
between the molecular BEC limit where $M=2m$ and the unitarity limit, where
$M$ becomes very large indicating that the dispersion relation is not
quadratic but rather linear as a  Bogoliubov phonon.
For example, 
the pair susceptibility $\chi$, which enters selfenergies,  is
calculated with different combinations of bare ($G_0$) and full ($G$)
Green's functions. With $\chi=GG$ \cite{Haussmann,Strinati} 
$M$ increases monotonically 
 as the unitarity limit is approached from the molecular BEC
limit, whereas with
$\chi=G_0G_0$ \cite{NSR,Randeria} $M$ first
decreases and then increases as $x\to 0_+$.
 With $\chi=G_0G$ \cite{Levin} $M$  first increase, then decrease
and finally increase again as $x\to 0_+$.
The critical temperature has the
opposite behavior as $M$ because $T_c\propto 1/M$.

A normal BEC has a linear Bogoliubov dispersion relation $E_q=qc_T$
for temperatures below $T^*=2\pi a_Mn/M$ corresponding to
$x\la (a_M/a)(T_c/T)$. For $a_M=0.62a$ and $T\sim T_c/2$ the quadratic
dispersion relation is thus limited to $x\ga 1$.
Similar results are found in
several crossover model \cite{Levin,Melwyn}, where 
the molecular gas can on the BEC side ($x\ga 1$) be
described approximately by a dispersion relation
$E_q=q^2/2M$, with an effective mass $M\simeq 2m$.

The corresponding entropy per particle is then given by the standard result
\bea \label{SBEC}
 \frac{S}{N} = \sigma\frac{3}{2} \left(\frac{T}{T_c}\right)^{3/2} \,,
\eea
where $\sigma=5\zeta(5/2)/3\zeta(3/2)=0.856$ and
the critical temperature is
$T_c=2\pi\hbar^2(n/2\zeta(3/2))^{2/3}/M=0.218E_F(2m/M)$.
The normal mass density is on the BEC side $\rho_n=\rho(T/T_c)^{3/2}$. 
From the entropy and mass densities on the BEC side we
obtain for the mixing term $c_2^2 = \sigma(T/M) \rho_s/\rho$ and
$c_S^2-c_T^2=\sigma (T/M)\rho_n/\rho$.

In the BEC limit we thus obtain the classical
result for the first sound speed
\bea \label{u1BEC}
 u_1=\sqrt{\sigma \frac{T}{M}+c_T^2} \,,
\eea
and the standard second sound speed 
$u_2^2=c_T^2\rho_s/\rho = 4\pi \hbar^2a_M\rho_s/M^3$, or
\bea \label{u2BEC}
  u_2=\sqrt{\frac{a_Mk_F}{6\pi} [1-(T/T_c)^{3/2}]} \,v_F \,.
\eea

As we approach the unitarity limit we observe that {\it avoided
crossing} takes place between the two sound modes caused by the mixing
term in Eq. (\ref{sss}). The mixing acts similar to level repulsion
between otherwise degenerate states in quantum mechanical
systems. Instead of crossing the two sound speeds avoid crossing by an
amount proportional to the mixing term, and first sound turns
continuously into second sound and visa versa.
As the avoided crossing occurs predominantly on the BEC side it can be
described by using the BEC mass densities and entropies in
Eq. (\ref{sss}).  The resulting sound modes are shown in Fig. 1 with
$c_T$ and $T_c$ from the Leggett crossover model which has $a_M=2a$.
For comparison the sound modes are calculated in Fig. 2 with $c_T$
from QMC \cite{Casulleras}, which has $a_M=0.62a$.  In both cases the
entropy of Eq. (\ref{SBEC}) with $M=2m$ is assumed.

\section{BCS limit}
The thermodynamic quantities and sound speeds in BCS limit have been 
calculated 
in \cite{sound} from the mean field BCS equations of Leggett \cite{Leggett}
\bea \label{gap}
  \frac{1}{g} = \sum_{\bf k} \left[
     \frac{1}{2E_k}-\frac{m}{\hbar^2k^2} -\frac{f_k}{E_k} \right] \,,
\eea
with coupling strength $g=-4\pi \hbar^2a/m$.
As usual $\varepsilon_k=\hbar^2k^2/2m-\mu$,
$E_k=\sqrt{\varepsilon_k^2+\Delta^2}$ and $\Delta(T)$ the gap. 
The thermal distribution function is
$f_k=(\exp(E_k/T)+1)^{-1}$.
With the equation for number density conservation
\bea \label{n}
  n = \sum_{\bf k}  \left[ 1-\frac{\varepsilon_k}{E_k} 
        +2\frac{\varepsilon_k}{E_k} f_k   \right] \,,
\eea
the gap and chemical potential can be calculated as function of
density, temperature and interaction strength.
At zero temperature
the last terms in Eqs. (\ref{gap}) and (\ref{n}) vanish and the gap is
\bea\label{D0}
 \Delta_0=\left(\frac{8}{e^2}\right)E_F\exp(\pi/2ak_F) \,.
\eea
 Gorkov \cite{Gorkov} included 
induced interactions, which reduce the gap to
\bea\label{DG}
 \Delta_0=(2/e)^{7/3}E_F\exp(\pi/2ak_F) \,. 
\eea
Induced interaction can be included
in the gap equation (\ref{gap}) by replacing 
$a^{-1}\to a^{-1}-2k_F\ln(4e)/3\pi$ on the left hand side.
Hereby, not only the gap is corrected but also the thermodynamic potential
\cite{sound}. In both cases $T_c=(\gamma/\pi)\Delta_0=0.567\Delta_0$.

The thermodynamic functions can be calculated from the
thermodynamic potential per volume $\Omega=-P$. We 
make the standard assumption that the Hartree-Fock terms in the superfluid
$\Omega_s$ and normal state $\Omega_n$ thermodynamic potentials
are the same.
The difference is then given in terms of the pairing coupling as
\bea\label{OS}
  \Omega_s &=& \Omega_n + \int_0^\Delta d\Delta' \Delta'^2 
         \frac{d(1/g)}{d\Delta'}  \,.
\eea
The first order temperature correction to the thermodynamic potential
in the normal phase is $\Omega_n=-N(0)\pi^2 T^2/3$, where
$N(0)=mk_F/2\pi^2$ is the level density.  

Inserting the coupling of Eq. (\ref{gap}) into the thermodynamic potential
(\ref{OS}) and again exploiting that $\Delta\ll\mu$,
it reduces to
\bea \label{OT}
   \Omega_s &=& -N(0)\left[ \Delta^2\left(\frac{1}{2}+
     \ln\frac{\Delta_0}{\Delta}\right)
       -4T\int_0^\infty d\varepsilon_k \ln(1-f_k) \right] \nonumber\\
   && 
\eea
The crossover model thus arrives at the standard expression for $\Omega_s$
and therefore also the standard
entropy density $S_s = -(\partial\Omega_s/\partial T)_{V,\mu}$,
and specific heat
$C_s = T(\partial S_s/\partial T)_{V,\mu}$ in superfluid phase.
In the normal phase $S_n=C_n=N(0)2\pi^2 T/3$.
 Near $T_c$:
$S_s/S_n(T)=1-(1+\xi)(1-T/T_c)$  and
$C_s/C_n(T_c)=\xi -3.77(1-T/T_c)$. 
Here, $\xi=1+12/7\zeta(3)\simeq2.43$ is the ratio of superfluid 
specific heat just below $T_c$ to the normal one just above $T_c$.
At low temperatures 
$S_s=N(0)\sqrt{2\pi\Delta_0^3/T}\exp(-\Delta_0/T)$ and
$C_s=N(0)\sqrt{2\pi\Delta_0^5/T^3}\exp(-\Delta_0/T)$.

Finally, we need the superfluid (London) density
\bea
   n_s &=& n\left(1+2\int^\infty_0 d\varepsilon_k \frac{df_k}{dE_k} \right)\,, 
\eea
where $n=n_s+n_n$ is the total density.
At low temperatures $n_s/n=1-\sqrt{2\pi\Delta_0/T}\exp(-\Delta_0/T)$
whereas  $n_s/n=2(1-T/T_c)$ near $T_c$.

We now have all the thermodynamic quantities available for calculating the
sound speeds in the BCS limit. Since thermal effects are small we find
that
$u_1=c_S\simeq c_T$. The second sound is a pure thermal wave with
velocity
\bea
  u_2^2 \simeq c_2^2= \frac{n_s}{n_n}\frac{S_s^2T}{mnC_s} \,.
\eea
The dependence on temperature and $x$ is
shown in Fig. 3 as well as in Figs. 1 and 2 for three different temperatures.

At low temperatures the second sound is linear in temperature
\bea
  u_2=\frac{\sqrt{3}}{2} \frac{T}{E_F} \, v_F
    \,,\quad\quad T\ll T_c \,.
\eea
Around $T\simeq 0.7T_c$
the second sound speed has a broad maximum of
\bea
u_2\simeq 0.53 \frac{T}{E_F} v_F \,,\,\quad T\sim 0.7T_c\,.
\eea 
Near the critical temperature $T_c-T\ll T_c$
\bea \label{u2BCS}
   u_2=\frac{\pi}{\sqrt{\xi}}\frac{T}{E_F}
       \sqrt{1-\frac{T}{T_c}}\, v_F   \,.
\eea
The characteristic $u_2\propto \sqrt{1-T/T_c}$ behavior near $T_c$
is the same as in the BEC limit, Eq. (\ref{u2BEC}), 
but with a different prefactor.

\vspace{-0.5cm}
\begin{figure}
\begin{center}
\psfig{file=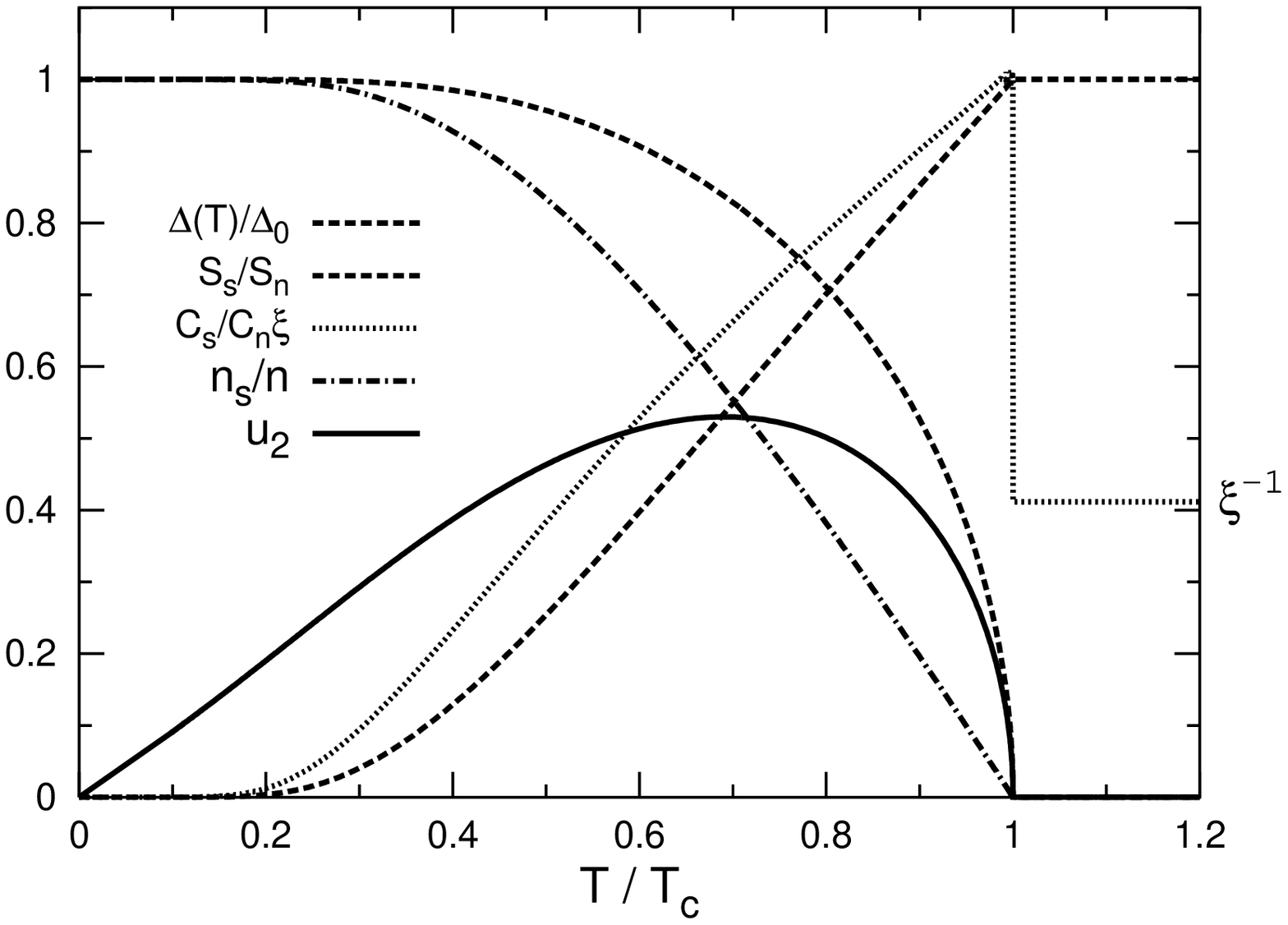,height=8.0cm,angle=-90}
\vspace{.2cm}
\begin{caption}
{Thermodynamic quantities for a superfluid as calculated in the BCS 
limit vs. $T/T_c$.
Shown are the gap $\Delta(T)/\Delta_0$, entropy $S_s(T)/S_n(T)$, 
specific heat $C_s(T)/C_n(T)\xi$, and superfluid density $n_s(T)/n$.
The second sound is plotted in units of $v_FT_c/E_F$.
}
\end{caption}
\end{center}
\label{f3}
\end{figure}
\vspace{-.2cm}

\section{Unitarity limit and Crossover} 

At crossover the entropy receives contributions from
thermal Fermi atoms and molecular bosons.  The entropies and sound
speeds reflect the underlying dispersion relation for low-lying
excitations.  On the BEC side for $T\ga T^*$
the quadratic dispersion relation
$E_q=\hbar^2q^2/2M$ leads to an entropy $S\propto
T^{3/2}$, whereas for $T\la T^*$ the
linear Bogoliubov phonon $E_q=\hbar qc_T$ yields 
$S\propto T^3$. 
On the BCS side the gap in the quasiparticle energy
leads to a more complicated entropy as discussed above.

Generally, for a power law dispersion relation $E_q=q^\alpha$ the
entropy and specific heat scale as $S\propto C_V\propto T^{3/\alpha}$ and
the normal mass density $\rho_n=\rho(T/T_c)^{5/\alpha-1}$.  Consequently
$c_S^2-c_T^2\propto T^{3/\alpha+1}$ and
$c_2^2\propto(\rho_s/\rho)T^{2-2/\alpha}$. The resulting
temperature dependence of the sound velocities follows from
Eq. (\ref{sss}).

Model calculations
\cite{Eagles,Leggett,NSR,Haussmann,Randeria,Levin,Strinati,Griffin,Holland,Melwyn}
find a smooth crossover of most thermodynamic quantities as function
of density and scattering length at zero and finite temperature except
at $T_c$, where superfluidity vanishes. We will therefore extrapolate
the sound modes from the BCS and BEC limits towards crossover such
that they are continuous around crossover and meet the sound speeds in
the unitarity limit, which now will be calculated independently.

Experimentally, the entropy has recently been measured and found to
scale as $S/N\simeq 20(T/T_F)^{1.53}$ in the unitarity limit at
temperatures below $T_c\simeq 0.27E_F$ \cite{entropy}.  In the
unitarity limit the entropy is independent of scattering length due to
universality \cite{Ho} as other thermodynamic quantities \cite{HH}.
Consequently, when $S/N\propto T^{3/2}$, the entropy density is
independent of density and is therefore the same in bulk as in a trap
at temperatures sufficiently low as compared to the Fermi temperature.
Both the measured $T_c$ and entropy are then slightly larger in the
unitarity limit than for the pure BEC of Eq. (\ref{SBEC}).  With these
adjustments of the parameters in Eq. (\ref{SBEC}), the sound
velocities are calculated in the unitarity limit as in the BEC limit
above as shown in Figs. 1 and 2 for $x=0$. To guide the eye the
unitarity limit point at $x=0$ is connected smoothly in the crossover
regime to the BEC limit, $x\ge1$, and the BCS limit, $x\le-1$.

The sound velocities in Figs. 1 and 2 are well behaved in the sense
that both the BEC and BCS limits extrapolate continuously towards the
unitarity limit.  On the BCS side first sound is basically given by
the isothermal sound speed, and the second sound increases
exponentially with $x$ as $T_c=(\gamma/\pi)\Delta$ as given by
Eqs. (\ref{D0}) and (\ref{DG}). Around the unitarity limit their
dependence on temperature and $x=1/k_Fa$ is more complicated because
the two sound modes mix and the temperature dependence of the entropy
is different from the BEC limit. On the BEC side the curves again
extrapolates continuously towards the BEC limit, however, with avoided
crossing of the two sound modes.

 It should be noted that the curves in Figs. 1 and 2 are plotted for
constant $T/T_c$. On the BEC side $T_c$ is almost constant from the
unitarity to the BEC limit, and the curves are therefore basically for
constant temperature. On the BCS side, however, $T_c$ decrease
exponentially as $T_c\propto\exp(\pi x/2)$ (see Eqs.(\ref{D0}) and
(\ref{DG})), and therefore the temperatures also decrease. If
the temperature were held fixed, superfluidity and the second sound would
thus vanishes below a certain $x$ where $T\ge T_c$.

The specific heat is a continuous function of temperature around $T_c$
only in the BEC limit. It is discontinuous in the BCS limit by a
factor $C_s/C_n=2.43$. Recent calculations and experimental fits
\cite{Levin} at $x=0.11$ find $C_s/C_n=2.51\pm0.05$ at $T_c$. These
experiments indicate that the discontinuity in the heat capacity has a
maximum crossing over from the BCS value towards zero in the BEC
limit.  $T_c$ also has a maximum at crossover in several crossover
models \cite{NSR,Randeria,Levin}.

 The second sound speed squared generally vanishes with the superfluid
density at the critical temperature, i.e.  $c_2\propto
\sqrt{T_c-T}$. This is observed in the BCS limit of
Eq. (\ref{u2BCS})), in the BEC limit of Eq. (\ref{u2BEC}) and also in
the unitarity limit \cite{Ho}.

The sound speeds are continuous as function of $x$ at crossover as
long as $T/T_c$ is fixed as shown in Figs. 1 and 2. As function of
temperature first sound depends only little on temperature whereas
second sound vanishes at $T_c$ and the specific heat is
discontinuous. Similar temperature dependency were found in
Refs. \cite{Griffin,Domanski} at crossover.

\section{Summary} 

The first and second sounds have been calculated in the molecular BEC, the BCS
and the unitarity limits. It has been argued that the
sound modes interpolate continuously between these limits, and that
avoided crossing takes place on the BEC side due to mixing.

The mixing has the consequence that both sound modes can be excited
and detected both as density and thermal waves in traps.  In contrast
the second sound is a pure thermal wave in the BCS and BEC Bogoliubov
limits. In finite traps the sound waves could be excited in the center
and propagate towards the surface so that the local density decreases.
Consequently, $|x|=1/|a|k_F$ increases and the sound velocities change
as shown in Figs. 1 and 2.

The second sound mode reveals superfluidity and the critical
temperature. Measurements of both sound modes constrain the equation
of state, entropies, specific heat, etc. and thereby crossover models,
which predict very different behavior of, e.g., the molecular
scattering length in the BEC limit and effective molecular mass,
critical temperature, etc. at crossover.
The crossover and mixing of the sound modes will also be possible to
study in mixtures of  bosons and fermions. A number of different
superfluids may also be created in
optical lattices \cite{Wang}
(with corresponding Mott transitions) and possibly also
supersolids.


\end{document}